\documentclass[letterpaper,11pt]{emulateapj}
\usepackage{epsfig}
\usepackage{natbib}
\usepackage{graphics}
\usepackage{graphicx}
\usepackage{multirow}
\usepackage{amssymb}
\usepackage{color}
\usepackage{boxedminipage}

\def \kms {{\rm km/s}}
\def \kmsMpc {{\rm km~s$^{-1}$~Mpc$^{-1}$}}

\shorttitle{Mid-IR TF Calibration of SNIa}
\shortauthors{Sorce et al.}

\begin{document}

\title{The Mid-Infrared Tully-Fisher Relation: Calibration of the SNIa Scale and H$_0$}

\author{Jenny G. Sorce$^{1}$}
\email{j.sorce@ipnl.in2p3.fr}
\author{R. Brent Tully$^{2}$}
\author{H\'el\`ene M. Courtois$^{1,2}$}
\affil{$^1$Universit\'e Claude Bernard Lyon I, Institut de Physique Nucleaire, Lyon, France} 
\affil{$^2$Institute for Astronomy, University of Hawaii, 2680 Woodlawn Drive, HI 96822, USA}

\begin{abstract}

This paper builds on a calibration of the SNIa absolute distance scale begun with a core of distances based on the correlation between galaxy rotation rates and optical $I_C$ band photometry.  This new work extends the calibration through the use of mid-infrared photometry acquired at $3.6 \mu$m with {\it Spitzer Space Telescope}.  The great virtue of the satellite observations is constancy of the photometry at a level better than 1\% across the sky.  The new calibration is based on 39 individual galaxies and 8 clusters that have been the sites of well observed SNIa.  The new $3.6 \mu$m calibration is not yet as extensively based as the $I_C$ band calibration but is already sufficient to justify a preliminary report.  Distances based on the mid-infrared photometry are $2\%$ greater in the mean than reported at $I_C$ band.  This difference is only marginally significant.  The $I_C$ band result is confirmed with only a small adjustment.  Incorporating a 1\% decrease in the LMC distance, the present study indicates H$_0 = 75.2 \pm3.0$~\kmsMpc.  

\end{abstract}


\section*{1. Introduction}

Type Ia supernovae (SNIa) have remarkable properties such as their high luminosities ($10^9\, L_{\odot}$) and their apparent homogeneous nature \citep{1995ApJ...438L..17R}.  \citet{1968AJ.....73.1021K} established the first Hubble diagram that suggested SNIa could be used as extragalactic distance indicators. Two decades later, \citet{1993ApJ...413L.105P} demonstrated the existence of a decline rate-absolute magnitude dependence for SNIa, validating that type Ia supernovae can act as standard candles. Work in subsequent years \citep{1995AJ....109....1H, 2007ApJ...659..122J, 2009ApJ...700.1097H, 2010ApJ...716..712A} has produced alternate descriptions of the correlations between the intrinsic luminosities of SNIa and the shapes of their light curves.

The properties of SNIa can be used to determine distances to galaxies at many hundreds of megaparsecs. At such distances, object are expected to have recessional velocities that individually differ from the mean by at most a few percent and collectively should define the cosmic expansion. Thanks to the great precision of SNIa distance estimates, high redshift SNIa revealed that the expansion of the universe is currently accelerating \citep{1998AJ....116.1009R, 1999ApJ...517..565P}. The SNIa method can provide the best estimate of the Hubble parameter once the zero point scale is set.  Independent distances are needed to the hosts of low redshift SNIa \citep{2009ApJ...699..539R, 2011ApJ...730..119R, 2010AJ....139..120F} to establish the absolute scale.

Our collaboration has recently contributed to the establishment of the SNIa scale \citep{2012ApJ...749..174C} primarily using constraints imposed by using distances acquired with the correlation between the luminosities and rotation rates of galaxies \citep{1977A&A....54..661T}, the Tully-Fisher relation (TFR).  Optical $I_C$ band luminosities were used in that study.   Now there is the opportunity to refine the calibration with the use of photometry at $3.6 \mu$m obtained with {\it Spitzer Space Telescope} \citep{2004ApJS..154....1W}.   The great advantage with Spitzer observations is photometric integrity to better than 1\% across the sky.  Additional advantages are minimal obscuration either within hosts or from our Galaxy, magnitude measures approximating total magnitudes because of low backgrounds, and fluxes dominated by light from old stars which presumptively correlates with galaxy mass.  By now, roughly 3000 galaxies have been observed with {\it Spitzer} and almost 1300 galaxies are being observed in the current cycle with our {\it Cosmic Flows with Spitzer} project (CFS\footnote{http://adsabs.harvard.edu/abs/2011sptz.prop80072T}).  Already, 39 galaxies have been observed that have hosted SNIa and are appropriate for an application of the TFR methodology. 

The present discussion will closely parallel the paper by \citet{2012ApJ...749..174C} with the important difference being the use of mid-infrared $[3.6]$ photometry in place of optical $I_C$ photometry.
We begin with a brief summary of the data that are available on the hosts of SNIa galaxies and the treatment given to obtain distances using the TFR in the mid-infrared. Distance measurements obtained via the TFR are individually uncertain. Averaging over a cluster provides a more robust distance so we include clusters in our analysis. Distances determined with the TFR enable us to set a zero point for the SNIa distance scale. Consideration of a large sample of SNIa in the redshift range $0.03 < z < 0.5$ leads us to an estimate of the Hubble Constant.

\section*{2. Data}

Three parameters are needed to obtain distances with the TFR: a luminosity, a measure of rotation, and an inclination to account for projection effects.  Our sample in this study is a subset of the sample used for the same purpose of a determination of SNIa host absolute luminosities by \citet{2012ApJ...749..174C}.  In the current paper we use the same information on rotation rates, from HI profile information, and inclinations, from optical band imaging.  The difference in this work is the replacement of $I_C$ luminosities with $[3.6]$ luminosities from observations using {\it Spitzer Space Telescope} IRAC channel 1.  Presently, not all the galaxies included in the $I_C$ band study have been satisfactorily observed with {\it Spitzer}. 
We have retrieved data from the {\it Spitzer Heritage Archive}  for 39 galaxies that have hosted SNIa from the list of 56 galaxies given by \citet{2012ApJ...749..174C}. The archival material comes from the programs {\it Spitzer Infrared Nearby Galaxies Survey}, SINGS, \citep{2005ApJ...633..857D,2007ApJ...655..863D}, {\it Spitzer Survey of Stellar Structure in Galaxies}, $S^{4}G$, \citep{2010PASP..122.1397S}, {\it Carnegie Hubble Program}, CHP, \citep{2011AJ....142..192F}, and our ongoing {\it Cosmic Flows with Spitzer} project, CFS.

	Photometry for these galaxies is carried out using a Spitzer-adapted version of Archangel \citep{2012PASA...29..174S} described in \citet{sorce2012a}. Luminosity corrections for IRAC channel 1 flux measurements were described in detail in \citet{sorce2012a}. Briefly, observed total magnitudes, $[3.6]$, must be corrected for extinction both within our Galaxy, $A_b^{[3.6]}$ \citep{1998ApJ...500..525S}, and within the hosts, $A_i^{[3.6]}$ \citep{1998AJ....115.2264T},  $k$-corrected, $A_k^{[3.6]}$ \citep{1968ApJ...154...21O, 2007ApJ...664..840H}, and receive an aperture correction, $A_a^{[3.6]}$ \citep{2005PASP..117..978R}. 
A fully corrected magnitude $[3.6]^{b,i,k,a}$ in the AB system for IRAC ch.1 {\it Spitzer} data is 
\begin{equation}
[3.6]^{b,i,k,a} = [3.6] - A_b^{[3.6]} - A_i^{[3.6]} - A_k^{[3.6]}+A_a^{[3.6]}
\end{equation}

\section*{3. Host Distances}

Our calibration of the TFR at $3.6 \mu$m is described in \citet{sorce2012b}.  The zero point is primarily set by Cepheids assuming a Large Magellanic Cloud (LMC) modulus $18.48 \pm 0.03$ (Freedman et al. 2012). 
The calibration used the correlation that assumes all errors are in the distance-independent line width parameter, the so-called `inverse' fit.  Corrections must be made to account for a small Malmquist bias effect with bias $b = -0.0065(\mu -31)^2$ where $\mu$ is the distance modulus.  Of greater importance is a color term. The calibration paper describes the tightening of the correlation that is provided by the adjustment to observed magnitude based on a color differential between the near-infrared $I_C$ band and the mid-infrared $[3.6]$ band
\begin{equation}
\Delta [3.6]^{color} = -0.36 - 0.47(I_C - [3.6])
\end{equation}
whence $C_{[3.6]} = [3.6]^{b,i,k,a} - \Delta [3.6]^{color}$.  Absolute color adjusted magnitudes $M_{C_{[3.6]}}$ are given by the equation
\begin{equation}
M_{C_{[3.6]}} = -20.34 - 9.13 ({\rm log} W_{mx}^{i} -2.5)
\end{equation}
where $W^i_{mx}$ is the HI line width, de-projected from inclination $i$ to edge-on, and adjusted to approximate twice the maximum rotation velocity of the galaxy \citep{2011MNRAS.414.2005C}. The rms scatter about this mid-IR version of the TFR is $\pm$0.42 mag in the calibration sample. A distance modulus is given by $\mu^c = C_{[3.6]} - M_{C_{[3.6]}} - b$.  These parameters are accumulated in Table~\ref{tbl:sn} for 39 galaxies that have hosted SNIa and been observed with {\it Spizer Space Telescope}.

Thirteen clusters were used to form the calibration template for the $[3.6]$ band TFR \citep{sorce2012b} so there is a good distance determination for each of these clusters. Suitably observed SNIa have been observed in 8 of these clusters.  Pertinent information is provided in Table \ref{tbl:clust}.  With the 2 nearest clusters (Virgo and Fornax) high quality distance measures are available from Cepheid and Surface Brightness Fluctuation observations and these measures contribute to (indeed, dominate) the values of the moduli in column 3 of the table.  The averaging over multiple contributions follows \citet{2012ApJ...749..174C}.  When there were more than one SNIa observed per galaxy or cluster, or more than one observation per SNIa, we take averaged SNIa modulus estimates \citep{2012ApJ...749..174C}.  The SNIa information is discussed in the next section.	

\section*{4. SNIa zero point scale and H$_0$}

\begin{figure}[h!]
\centering
\includegraphics[scale=0.44]{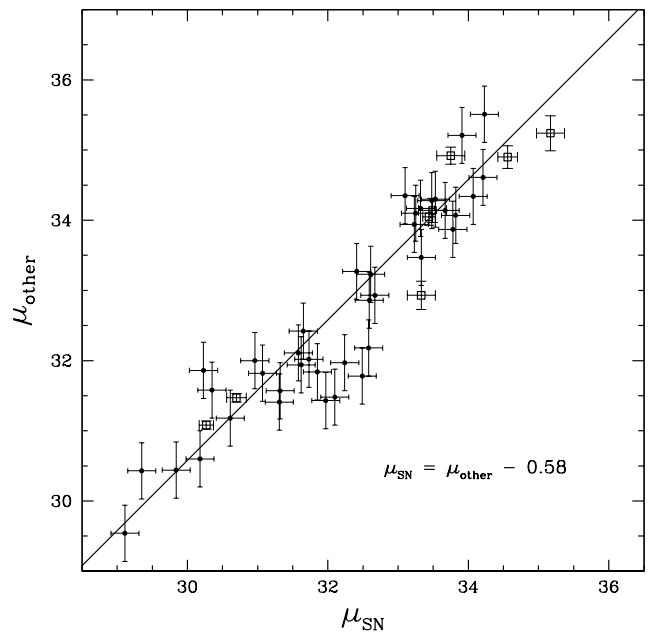}
\caption{Comparison between moduli derived with SNIa and with `other' methods: the TFR, with Cepheid and Surface Brightness Fluctuation supplements. The comparisons include 39 individual galaxies with TFR measurements (filled points) and 8 clusters (open squares).  The straight line is a weighted fit to the 39 galaxies with TFR distances and 6 of the 8 clusters.}
\label{comparison}
\end{figure}

\citet{2012ApJ...749..174C} discuss the accumulation of a sample of SNIa from 5 sources \citep{2006ApJ...647..501P, 2007ApJ...659..122J, 2009ApJ...700.1097H, 2010AJ....139..120F, 2010ApJ...716..712A} with scale shifts as appropriate to match the scale of the last of these sources, a compilation referred to as UNION2.  Relevant distance moduli are gathered from these 5 sources and recorded in Table~\ref{tbl:clust} with averaging in the case of clusters with multiple recorded SNIa events.  Moduli drawn from Tables 1 and 2 are compared in Figure~\ref{comparison}.  

The straight line in this figure is a fit, assuming slope unity, to the 39 individual galaxies each with weight 1 and 6 clusters each with weight 9.  The locations of two clusters are deviant (Centaurus at $5\sigma$ under the fit in Fig.~\ref{comparison} and A1367 at $3\sigma$ over the fit). These two clusters were deviant and rejected from the optical SNIa calibration \citep{2012ApJ...749..174C} and for consistency in the comparison are again rejected from the fit.  The offset between the newly determined distance moduli (other) and the SNIa moduli on the UNION2 scale is $\mu_{other} - \mu_{SN} = 0.58$. The comparable fit with $I_C$ band material was shown in Figure~5 of  \citet{2012ApJ...749..174C}.  The offset in that earlier case was 0.56. The current calibration increases distances by 1\% and reduce H$_0$ by 1\%.

\begin{figure}[h!]
\centering
\includegraphics[scale=0.44]{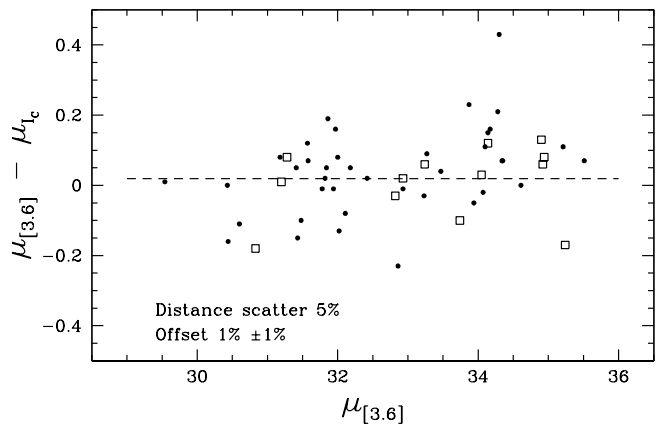}
\caption{Differences in TFR distance moduli measured at [3.6] and at $I_C$ plotted against the [3.6] moduli. Filled points: the 39 galaxies that have hosted SNIa; open squares: the 13 TFR template calibrator clusters.}
\label{LmI}
\end{figure}

The galaxies observed to date with Spitzer are only a subset of those discussed in the $I_C$ band calibration paper.  It is instructive to compare results using only identical galaxies and clusters rather than using the ensemble of available samples as was done above.  Figure~\ref{LmI} compares distance moduli measured alternatively with mid-IR [3.6] photometry with {\it Spitzer} and optical $I_C$ photometry observed from the ground, using the same line width and inclination parameters.  The comparison involves the 13 clusters used to establish the TFR template at $I_C$ \citep{2012ApJ...749...78T} and [3.6] (Sorce et al. 2012b) and the 39 individual galaxies that have hosted SNIa (\citet{2012ApJ...749..174C} and this paper).  With the individual SNIa hosts there is a hint of an increase in the difference between moduli for the more distant cases but the trend is not statistically significant.  No such trend is seen with the clusters.  Overall the [3.6] moduli are greater than the $I_C$ moduli by $0.02 \pm 0.02$ mag.  The difference of 1\% in distance is not statistically significant.  It is to be noted, though, that the new mid-IR calibration is tied to a distance to the Large Magellanic Cloud that is 1\% closer than previously assumed (Sorce et al. 2012b).  With a common choice of LMC distance, the [3.6] band distances are 2\% greater than those at $I_C$ band.

The final calibration of the SNIa distance scale in the $I_C$ band analysis of \citet{2012ApJ...749..174C} lead to the determination of the Hubble Constant shown in their Figure~8.  It is based on a fit over the redshift range $0.03 <  z < 0.5$ to the UNION2 sample \citep{2010ApJ...716..712A}, with cosmological parameters $\Omega_m = 0.27$, and $\Omega_{\Lambda} = 0.73$.  The result obtained in that paper was H$_0 = 75.9 \pm 3.8$~\kmsMpc.  In the present work, distances are decreased 1\% due to a revised Large Magellanic Cloud modulus and increased 2\% with the switch from optical $I_C$ to mid-IR $[3.6]$ magnitudes.
The present calibration is in statistical agreement with the earlier work though formally gives a result $1\%$ lower. 
An error budget was discussed by \cite{2012ApJ...749..174C}.  Uncertainties are reduced with this new work in two respects.  First, there is increased confidence in the absolute scale set by the distance to the LMC (Freedman et al. 2012).  Second, the mid-IR calibration of the TFR (Sorce et al. 2012b) removes latent concerns about possible photometric differences in different parts of the sky.  These two improvements warrant a decrease in our error estimate from 5\% to 4\%.  
Our best estimate for the Hubble Constant is now H$_0 = 75.2 \pm3.0$~\kmsMpc.

\section*{5. Conclusions}

The new mid-infrared TFR calibration of the SNIa distance scale leads to a result for the Hubble Constant that is not significantly different from the earlier optical TFR calibration. 
The earlier calibration made use of a considerably larger collection of material.  Besides using over 50\% more individual TFR galaxies, it gave consideration to 61 groups or clusters hosting SNIa with distances not only from the TFR but also with Cepheid, surface brightness fluctuation, and fundamental plane measurements.  Nevertheless we contend that the present confirming work has value because it puts to rest a concern with the optical study.  The optical photometry was acquired by a diverse community of observers on several telescopes with a variety of detectors and filters and subject to the vagaries associated with ground-based observations.  This new mid-IR photometry is being acquired with a single observing configuration in space advertising photometric consistency across the sky to better than 1\%.  In the fullness of time it can be anticipated that the mid-IR calibration of the distance scale will be made more robust with linkages to SNIa involving several hundred galaxies.  The paper describing the calibration of the mid-IR TFR had already lead to a preliminary determine of the Hubble Constant of H$_0 = 74 \pm4$~\kmsMpc \citep{sorce2012b}.  The present study extends the calibration to distances where peculiar velocities should have negligible impact and we find H$_0 = 75.2 \pm3.0$~\kmsMpc.

\bigskip

We especially thank our {\it Cosmic Flows with Spitzer} collaborators Wendy Freedman, Tom Jarrett, Barry Madore, Eric Persson, Mark Seibert, and Ed Shaya although most of the data used in this paper comes from the {\it Spitzer} archive.  We are indebted to James Shombert for the development and support of the Archangel photometry package.  We thank Kartik Sheth for discussions regarding {\it Spitzer} photometry.  NASA through the {\it Spitzer Science Center} provides support for {\it Cosmic Flows with Spitzer}  cycle 8 program 80072.  
RBT receives support for aspects of this program from the US National Science Foundation with award AST-0908846.

\clearpage

\begin{table}[h!]
\caption{Properties of individual SNIa galaxies}
\begin{tabular}{|l|r|l|r|c|r|r|c|c|c|}
\hline
Name\tablenotemark{a} & PGC\tablenotemark{b} & SNIa\tablenotemark{c} & $V_{CMB}$\tablenotemark{d} & $W_{mx}^i$\tablenotemark{e} & $[3.6]^{b,i,k,a}$ \tablenotemark{f} & $C_{[3.6]}$\tablenotemark{g} & $M_{C_{[3.6]}}$\tablenotemark{h} & $\mu_{TF}$\tablenotemark{i} & $\mu_{SN}$\tablenotemark{j} \\
\hline

UGC00646 & 3773 & 1998ef & 5011 & 389 & 12.89 & 12.88 & -21.16 &  34.10 & 33.25 \\
PGC005341 & 5341 & 1998dm & 1663 & 236 & 12.91 & 12.79 & -19.18 & 31.97 & 32.24 \\   
NGC0673 & 6624 & 1996bo & 4898 & 445 & 12.04 & 12.19 & -21.69 & 33.94 & 33.23 \\
 NGC0958 & 9560 & 2005A & 5501 & 584 & 11.09 & 11.24 & -22.77 & 34.07 & 33.82 \\
ESO300-9 & 11606 & 1992bc & 5918 & 323 & 14.69 & 14.67 & -20.42 & 35.21 & 33.91 \\
NGC1148 & 13727 & 2001el & 1092 & 386 & 10.05 & 10.05 & -21.13 & 31.18 & 30.61 \\
UGC03329 & 17509 & 1999ek & 5277 & 525 & 12.13 & 11.89 & -22.35 & 34.30 & 33.53 \\
UGC03375 & 18089 & 2011gc & 5792 & 535 & 11.82 & 11.78 & -22.42 & 34.28 & 33.48 \\
PGC018373 & 18373 & 2003kf & 2295 & 234 & 12.72 & 12.69 & -19.15 & 31.84 & 31.85 \\
UGC03432 & 18747 & 1996bv & 5015 & 289 & 14.20 & 14.12 & -19.98 & 34.17 & 33.32 \\
UGC03576 & 19788 & 1998ec & 6013 & 393 & 13.03 & 13.07 & -21.20 & 34.34 & 34.07 \\
UGC03370 & 20513 & 2000fa & 6525 & 371 & 13.48 & 13.56 & -20.97 & 34.61 & 34.21\\
UGC03845 & 21020 & 1997do & 3136 & 257 & 13.35 & 13.39 & -19.52 & 32.93 & 32.67 \\
NGC2841& 26512 & 1999by & 804 & 650 & 8.68 & 8.66 & -23.20 & 31.86 & 30.23 \\
NGC3021 & 28357 & 1995al & 1797 & 303 & 11.68 & 11.84 & -20.17 & 32.02 & 31.73 \\
NGC3294 & 31428 & 1992G & 1831 & 431 & 10.77 & 10.84 & -21.57 & 32.42 & 31.65 \\
NGC3368 & 32192 & 1998bu & 1231 & 428 & 8.80 & 8.89 & -21.54 & 30.43 & 29.35 \\
NGC3370 & 32207 & 1994ae & 1609 & 312 & 11.68 & 11.81 & -20.29 & 32.11 & 31.58 \\
NGC3627 & 34695 & 1989b & 1061 & 385 & 8.33 & 8.40 & -21.12 & 29.54 & 29.11 \\
NGC3663 & 35006 & 2006ax & 5396 & 443 & 12.44 & 12.41 & -21.68 & 34.14 & 33.67 \\
NGC3672 & 35088 & 2007bm & 2223 & 399 & 10.59 & 10.68 & -21.26 & 31.94 & 31.62 \\
NGC4501 & 41517 & 1999cl & 2601 & 570 & 8.84 & 8.90 & -22.68 & 31.58 & 30.35 \\
NGC4527 & 41789 & 1991T & 2072 & 362 & 9.34 & 9.56 & -20.88 & 30.44 & 29.84 \\
NGC4536 & 41823 & 1981B & 2144 & 341 & 9.81 & 9.96 & -20.64 & 30.60 & 30.18 \\
NGC4639 & 42741 & 1990N & 1308 & 336 & 11.18 & 11.23 & -20.58 & 31.82 & 31.07 \\
NGC4680 & 43118 & 1997bp & 2824 & 237 & 12.09 & 12.23 & -19.20 & 31.43 & 31.97 \\
NGC4679 & 43170 & 2001cz & 4935 & 427 & 11.83 & 11.90 & -21.53 & 33.47 & 33.33 \\
NGC5005 & 45749 & 1996ai & 1178 & 601 & 9.05 & 9.11 & -22.89 & 32.00 & 30.96 \\
ESO576-040 & 46574 & 1997br & 2385 & 170 & 13.82 & 13.69 & -17.88 & 31.57 & 31.32 \\
PGC47514 & 47514 & 2007ca & 4517 & 285 & 14.03 & 13.89 & -19.93 & 33.87 & 33.78 \\
NGC5584 & 51344 & 2007af & 1890 & 267 & 11.75 & 11.74 & -19.67 & 31.41 & 31.31 \\
IC1151 & 56537 & 1991M & 2274 & 242 & 12.91 & 12.88 & -19.28 & 32.18 & 32.58 \\
NGC6063 & 57205 & 1999ac & 2950 & 308 & 13.06 & 13.01 & -20.24 & 33.27 & 32.41 \\
UGC10738 & 59769 & 2001cp & 6726 & 585 & 12.52 & 12.61 & -22.78 & 35.51 & 34.23 \\
UGC10743 & 59782 & 2002er & 2574 & 206 & 12.74 & 12.83 & -18.64 & 31.48 & 32.10\\
NGC6962 & 65375 & 2002ha & 3936 & 633 & 11.11 & 11.19 & -23.09 & 34.35 & 33.10 \\
IC5179 & 68455 & 1999ee & 3158 & 444 & 10.86 & 11.15 & -21.69 & 32.86 & 32.59 \\
NGC7329 & 69453 & 2006bh & 3143 & 461 & 11.24 & 11.36 & -21.83 & 33.23 & 32.61 \\
NGC7448 & 70213 & 1997dt & 1838 & 316 & 11.37 & 11.44 & -20.34 & 31.78 & 32.49 \\
\hline
\end{tabular}
\tablenotemark{a}{Common name}\\
\tablenotemark{b}{PGC name}\\
\tablenotemark{c}{SNIa identification}\\
\tablenotemark{d}{Mean velocity of host galaxy with respect to the CMB, \kms}\\
\tablenotemark{e}{Corrected rotation rate parameter corresponding to twice maximum velocity, \kms}\\
\tablenotemark{f}{Corrected $3.6 \mu$m magnitude in the AB system, mag}\\
\tablenotemark{g}{Color adjusted magnitude, mag}\\
\tablenotemark{h}{Absolute color adjusted magnitude, mag}\\
\tablenotemark{i}{TFR distance modulus corrected for bias, mag}\\
\tablenotemark{j}{SNIa distance modulus, mag}
\label{tbl:sn}
\end{table}

\clearpage

\begin{table}[h!]
\caption{Properties of clusters with SNIa}
\begin{tabular}{|l|c|c|c|c|c|c|}
\hline
Cluster\tablenotemark{a} & $V_{CMB}$\tablenotemark{b} & $\mu_{other}$\tablenotemark{c} & No. CFS\tablenotemark{d} & $\mu_{SN}$\tablenotemark{e} & SNIa names\tablenotemark{f} \\
\hline
Virgo   & 1410 & 31.08 $\pm$  0.06 & 24 & 30.27 $\pm$ 0.10 & 1991bg, 1994D, 1999cl, 2006X  \\
Fornax & 1484  & 31.47 $\pm$  0.06 & 15 & 30.70 $\pm$ 0.14 & 1980N, 1992A\\
Cen30  & 3679  & 32.93 $\pm$ 0.20 & 11 & 33.33 $\pm$ 0.20 & 2001cz\\
Pisces  & 4779  & 34.05 $\pm$ 0.11 & 23 & 33.44 $\pm$ 0.09 & 1998ef, 1999ej, 2000dk, 2001en, 2006td\\
Cancer & 4940  & 34.14 $\pm$ 0.13 & 11 & 33.49 $\pm$ 0.20 & 1999aa \\
Coma  & 7194  & 34.90 $\pm$ 0.13 & 16 & 34.56 $\pm$ 0.14 & 2006cg, 2007bz\\
A1367 & 6923  & 34.92 $\pm$ 0.14 & 19 & 33.75 $\pm$ 0.20 & 2007ci\\
A2634/66 & 8381  & 35.24 $\pm$ 0.13  & 18 & 35.17 $\pm$ 0.20 & 1997dg\\
\hline
\end{tabular}
\tablenotemark{a}{Cluster name}\\
\tablenotemark{b}{Mean velocity of the cluster with respect to the CMB, \kms}\\
\tablenotemark{c}{TFR distance modulus corrected for bias (Virgo and Fornax are special cases discussed in text), mag}\\
\tablenotemark{d}{Number of galaxies used in the TFR calibration}\\
\tablenotemark{e}{SNIa distance modulus, mag}\\
\tablenotemark{f}{SNIa identifications}

\label{tbl:clust}
\end{table}

%%%%%%%%%%%%%%%%%%%%%%%%%%%%%%%%%%%%%%%%%%%%%%%%%%%%%%%%%%%%%%%%%%%%%%%%%
%%%%%%%%%%%%%%%%%%%%%%%%%%%%%%%%%%%%%%%%%%%%%%%%%%%%%%%%%%%%%%%%%%%%%%%%%
\clearpage

\bibliographystyle{ApJ}

\bibliography{bibli3}

\end{document}